# Discussing politics with a chatbot: How political beliefs shape prompt formulation for generative AI

*Victoria Vziatysheva[1,*] & Reinhold Kliegl[2]*

[1] Institute of Communication and Media Studies, University of Bern, Bern, Switzerland

[2] University of Potsdam, Potsdam, Germany

[*] Corresponding author: Victoria Vziatysheva. Email: victoria.vziatysheva@unibe.ch

## Abstract

By generating comprehensive and highly personalized answers to user prompts, AI chatbots such as ChatGPT and Gemini can serve as alternatives to search engines and substantially transform online information seeking. However, little research has examined how users interact with these tools to find information, particularly about politics. In this experiment, we examine how Swiss citizens (N = 563) formulate prompts to search for migration-related issues using AI chatbots and whether this process is affected by selective exposure. We find evidence of attitude-consistent prompting, whereby participants who were more critical of migration were more likely to formulate prompts with an anti-migration stance. This effect was stronger for a broad topic (migration in general) than for a specific one (anti-migration popular initiative put to a vote in Switzerland). However, exposure to an attitude-consistent AI response did not further amplify selectivity in prompting. We also did not find evidence that either AI attitudes or thinking styles moderated selectivity in prompting. These findings contribute to the understanding of selective exposure via AI chatbots and the use of generative AI for political information-seeking.

## Introduction

The rise of generative artificial intelligence (AI) has been rapidly transforming information-seeking practices. AI chatbots powered by Large Language Models (LLMs) and Multimodal Large Language Models (MLLMs)—such as ChatGPT, Claude, or Gemini—can quickly generate responses to user questions and access other digital sources. While generative AI is used for a variety of purposes, including text editing, making summaries, and brainstorming new ideas, information retrieval remains one of the primary motivations for employing these tools (Vziatysheva, 2026a). As search engines and social media integrate chatbot features into their interfaces, generative AI becomes an increasingly integral part of information seeking.

At the same time, generative AI offers features distinctive from those of traditional search engines. Instead of retrieving a selection of websites based on keywords, AI chatbots can generate natural language and "engage in natural conversational interactions with humans" (Khatri et al., 2018), making them highly anthropomorphic. Anthropomorphism, or attributing "humanlike characteristics, motivations, intentions, and emotions" to nonhuman agents (Epley et al., 2007), can, in turn, increase trust in AI systems (Cheng et al., 2022; Glikson & Woolley, 2020) and positive perceptions of them (Williams, 2026). Furthermore, this mode of information delivery may reduce users' need to visit other websites (Kaiser et al., 2025), therefore, potentially limiting the perspectives individuals are exposed to, including those in the political domain.

AI chatbots' performance can considerably vary depending on the prompts—questions and commands that users submit to the chatbots. On the one hand, better prompts improve the quality of the output (Marvin et al., 2024). On the other hand, the prompt formulation can provoke sycophantic behavior when an AI chatbot, aiming at users' approval, □re-emptively confirms their beliefs (M. Sharma et al., 2025). This may be particularly concerning in the context of political information seeking, since AI chatbots, being highly persuasive, can influence voters' beliefs (Lin et al., 2025) and, due to their agreeableness and personalization, reinforce individual opinions (Rathje et al., 2025). All this makes the choice of prompts a central aspect of users' information-seeking behavior via AI chatbots.

While a range of studies explore the effectiveness of different prompting strategies and their effects on output (Della Porta et al., 2025; Sawalha et al., 2024), fewer studies investigate how users formulate their prompts, especially about political issues. Yet, this aspect of interactions with AI chatbots is crucial, as ideologically slanted prompts can lead to ideologically slanted outputs (Bianchi et al., 2026). This study experimentally examines how users formulate prompts to seek political information via AI chatbots, drawing on selective exposure theory, which posits that people select information consistent with their beliefs (Stroud, 2017).

Specifically, we test the hypotheses that users would prefer prompts consistent with their beliefs and that this effect can be amplified by the search topic and exposure to attitude-consistent AI responses. Furthermore, we test whether thinking styles (effortful, intuitive, closed- and open-minded) and AI attitudes (trust in AI chatbots and perceived AI competence) moderate user preference for attitude-consistent prompts. We are studying this in the context of migration-related issues, information about which people often evaluate in line with their political ideology (e.g., Lind et al., 2022). This study is conducted in Switzerland, a semi-direct democracy and one of the early adopters of AI (Sajadieh et al., 2026), both of which make the use of AI chatbots highly relevant to citizens' political decision-making.

## Literature review

### Selective exposure to information online

Selectivity can be broadly understood as paying selective attention and exposing oneself to signals in the environment (Donsbach & Mothes, 2017). While selective exposure to information can take various forms—for example, prioritizing entertainment content over news or one topic over the other (Stroud, 2017)—communication research largely focuses on selective exposure as a preference for information consistent with one's attitudes, values, and opinions (Freedman & Sears, 1965; Hastall, 2020), which is also how we use this concept in the present study.

Whether people seek out attitude-consistent information and to what degree depends on the context. In their meta-analysis of factors affecting selective exposure, Hart et al. (2009) distinguish between accuracy and defense motivations driving information choices. While the former can reduce congeniality bias (for example, when uncongenial information is useful for decision-making), the latter can, on the contrary, amplify preference for attitude-consistent information (for example, when the issue is highly relevant to one's values). Migration can be considered one of the issues activating defense motivations, as attitudes towards it are largely related to values and social identity (Davidov & Meuleman, 2012; Konings et al., 2023).

Selectivity can occur at different stages of information processing—from search to perception and retention (Hastall, 2020). In an experiment simulating online search, Knobloch-Westerwick et al. (2015) found that users spent more time engaging with search results that aligned with their beliefs. Similarly, an experiment by Garrett (2009) showed that when participants were exposed to different online news stories, they preferred attitude-consistent messages. Yet, studies analyzing actual search histories paint a more complex picture (as compared to experimental settings). For instance, in a study of user browsing behavior during the Israeli election, Dvir-Gvirsman et al. (2016) found that although exposure to like-minded content was more frequent than cross-cutting exposure, users generally had diverse news diets. Using a web-tracking dataset in Spain, Cardenal et al. (2019) demonstrated that accessing news via Google reduced selective exposure (compared with direct navigation).

Importantly, these studies, as well as a large body of research on selective exposure, focus predominantly on preferences for different messages and sources. However, fewer studies examine the earlier stages of online search—namely, how users phrase their search requests to locate information of interest, even before they are exposed to different options. In this study, we are interested in the step preceding information selection—particularly, the formulation of prompts for AI chatbots.

## Selectivity in search requests

Existing research on how users formulate search requests across different platforms remains fragmented. A large body of studies examines search queries from an information retrieval perspective. These studies, for instance, offer taxonomies of search requests based on user intent and corresponding informational needs (Broder, 2002; Lichtenegger et al., 2026). Others evaluate search queries as part of user search strategies or the ability to successfully find information (Ford et al., 2005; Sanchiz et al., 2017).

Fewer studies examined search requests through the lens of selective exposure. Focusing primarily on search engine queries, this research yields mixed evidence on how individual (political) beliefs influence query selection. Several studies, like the present one, examined how Swiss citizens search for information about popular initiatives and whether their search requests align with their voting intentions (Blassnig et al., 2023; Vziatysheva, 2025; Vziatysheva et al., 2026). Using self-reported data, neither study found evidence that search query stance was related to how a person intended to vote on a specific proposal. However, one study found selectivity in search queries collected through donated search history (Blassnig et al., 2023), and another demonstrated that queries about an environmental popular initiative were partially driven by climate beliefs (Vziatysheva, 2025).

Other studies examined search query selection in the context of broader and more contentious topics. Looking at issues such as immigration, abortion, and climate change, Ekström et al. (2024) and Slechten et al. (2022) found that search query selection can indeed be driven by attitudes regarding a specific topic. Similarly, van Hoof et al. (2024) found that attitudes towards immigration influence the formulation of search queries. Such differences can be explained by both differing information needs (e.g., interest in one aspect of the topic over another) and linguistic styles, which can reflect one's political ideology (Sterling et al., 2020).

## AI chatbots and information seeking

The rise of AI chatbots is significantly changing information-seeking practices, prompting a revisiting of the concept of selective exposure. While earlier selective exposure research largely focused on how individuals select sources and messages, this perspective may become less relevant in the context of interactions with AI chatbots. Firstly, users typically have limited opportunities to select information (beyond choosing an AI chatbot itself), as chatbots often provide complete answers to users' questions rather than a list of sources. Second, even when primary sources are referenced in AI-generated output, research shows that, compared with search engines, users are less likely to visit them (Kaiser et al., 2025). Third, interactions with AI typically require longer, more precise prompts than search engine queries (Trippas et al., 2024). In this sense, the process of obtaining the desired information is largely shaped by the prompt formulation rather than selecting one search result over another. Furthermore, the acquisition of information shifts from selection to co-creation, as a user is able to guide the generation of AI responses by asking additional questions, providing feedback, correcting, or requesting modifications to chatbots' outputs. Thus, user-driven selectivity is likely to occur in prompting—both during initial question-posing and in subsequent reactions to AI-generated responses.

Furthermore, specific features of AI chatbots—such as a high level of personalization, sycophancy, and anthropomorphism—affect how users interact with these systems. Due to chatbots' ability to engage in human-like conversations, users are more likely to use personified language or disclose personal information (Croes et al., 2023; Lee et al., 2024). Some use these tools for advice, emotional support, or companionship (Denecke et al., 2021;

Kostka & Zhou, 2026; Manoli et al., 2026), making interactions with AI more personal. In the political context, this may affect both the form (e.g., language used) and the content of prompts (e.g., seeking advice or reassurance).

One of the few studies analyzing prompt formulation found evidence of attitude-confirming information seeking, which was especially prominent when a chatbot was fine-tuned to match users' views on the issue (N. Sharma et al., 2024). Furthermore, there is evidence that chatbots' predictive sycophancy (or tendency to reflect users' viewpoints) increases when models can accurately infer users' political views (Jain et al., 2026). Thus, we can assume that prompts—serving as cues for one's political views—may contribute to more sycophantic AI responses on political issues. Considering the growing use of chatbots for political information, particularly in the electoral context (Luettgau et al., 2026), it is necessary to further investigate whether, and under what conditions, selective exposure patterns occur in prompts for AI chatbots.

## Pre-registered hypotheses

The main assumption of this study was that users formulate prompts aligned with their beliefs. We examine this in the context of migration-related information and hence pre-registered the following hypothesis:

*H1: The more negative users' beliefs regarding migration are, the more likely they will be to formulate negative prompts about migration-related issues, and vice versa.*

As discussed earlier, selective exposure may depend on the topic, with issues related to one's values and identity more likely to trigger defense motivations. In this study, we compare selective exposure in the context of a broader debate around migration and a more specific migration-related policy. Thus, we formulated the following hypothesis:

*H2: Users are more likely to formulate prompts consistent with their beliefs when searching for a broad topic (migration) rather than for a specific policy (migration-related initiative).*

Finally, research showed that exposure to pro- or counterattitudinal information can increase polarization (Wojcieszak & Garrett, 2018) and that sequential exposure to information results in a stronger confirmation bias (Jonas et al., 2001). Based on that, we expected that exposure to an AI response confirming or contradicting one's beliefs would affect subsequent prompting.

*H3: Users are more likely to formulate attitude-consistent prompts after exposure to an AI response aligned with their beliefs.*

## Exploratory research questions

We further explored whether other individual factors can moderate selectivity in prompting. First, we tested the effect of AI attitudes and knowledge, particularly trust in AI chatbots and perceived AI competence. It is known that both AI trust and various aspects of AI literacy can influence users' perceptions of, or interactions with, AI systems—such as the intention to use the technology, its acceptance, or engagement (Bach et al., 2024; Schiavo et al., 2024; Yilmaz Soylu et al., 2025). Thus, we posed the following question:

*RQ1: Do trust in AI chatbots and perceived AI competence moderate the relationship between migration beliefs and prompt stance?*

Second, we examined whether specific thinking styles—intuitive vs. effortful, closed-minded vs. actively open-minded (Newton et al., 2024)—can influence users' propensity to formulate attitude-consistent prompts. Predispositions to different thinking styles are often discussed within dual-process theories, which juxtapose two types of reasoning: Type 1, responsible for quicker, more intuitive information processing, and Type 2, responsible for more effortful, analytical reasoning (Evans & Stanovich, 2013). While these factors have not yet been studied specifically in relation to prompting, research has examined how cognitive factors affect the processing of attitude-consistent information. For instance, scholars suggest that actively open-minded thinking may reduce politically motivated reasoning (Stenhouse et al., 2018), whereas effortful thinking may, by contrast, amplify it (Kahan, 2013). Furthermore, closed-mindedness is associated with a higher preference for congenial information (Hart et al., 2009). Thus, we asked the following research question:

*RQ2: Do thinking styles moderate the relationship between migration beliefs and prompt stance?*

# Methods

## Design

This study is based on an online survey experiment with a 2 (Topic) x 3 (AI response; further, AIR) x 2 (Time) design. Topic and AIR are between-subject factors, and Time (pretest/posttest) is a within-subject factor. Preregistration of the hypotheses and the analysis are available under: https://osf.io/vb8rz/overview?view_only=e4663cb42cf349aaaf23d91d5e074a87. The study design is shown in Figure 1.

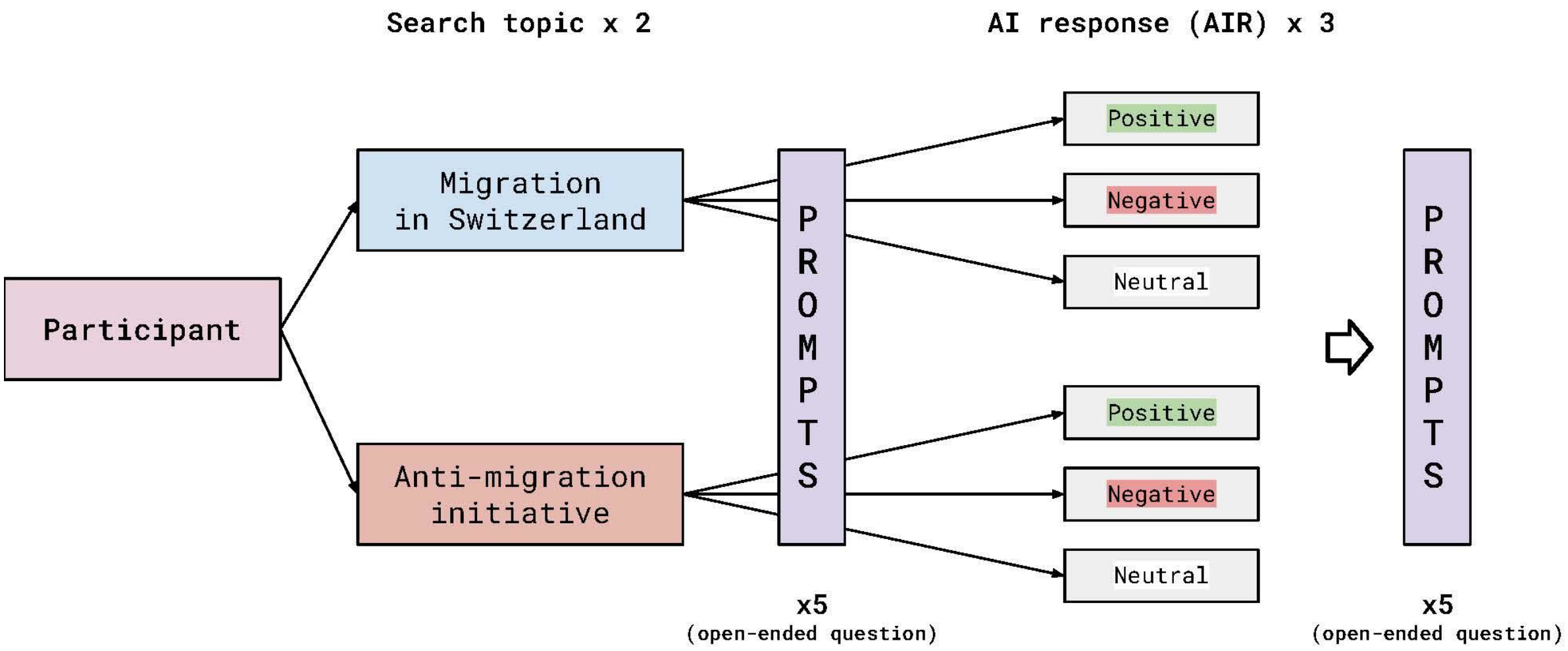

**Figure 1. Experimental design**

First, participants answered questions about their AI use, political leaning, and migration beliefs. Then, they were randomly assigned to one of the topics (broad or specific) and asked to formulate five prompts for an AI chatbot to find more information about that topic. Prompts were collected as open-ended questions. Subsequently, participants were presented with an AI response in one of three conditions (positive/negative/control). After that, they were asked to provide five additional prompts they would use once they received this information. At the end, participants answered questions measuring their thinking styles. Furthermore, we included two manipulation checks: (1) factual—testing whether participants read the stimuli (i.e., asking what it was about), (2) subjective—testing the strength of the stimuli (i.e., asking whether a stimulus framed migration in a positive, negative, or neutral way).

The experiment was programmed using Qualtrics. The data was collected in August 2025 with the help of the market research company Kantar. The study was approved by the Ethics Committee of the Faculty of Business and Social Sciences of the University of Bern (serial number 312025). Another set of hypotheses concerning the perception of AI responses was tested using the same data and is reported in [anonymized]. All supplementary materials—including stimuli, the questionnaire, and data analysis—are available in a public repository: https://osf.io/vb8rz/overview?view_only=e4663cb42cf349aaaf23d91d5e074a87.

## Sample

The sample included 563 participants (279 women, 284 men, M age = 44.35). All participants were German-speaking Swiss nationals, at least 18 years old, and had previously used an AI chatbot at least once. The sample was collected according to the gender and age quotas representative of the Swiss population, with a slight underrepresentation of people over 55.

Initially, 852 complete responses were collected; however, participants who failed the factual manipulation check (that is, did not correctly answer what the stimulus was about), did not answer any of the migration beliefs items (the key independent variable), or did not provide a single meaningful prompt (the dependent variable), were excluded.

## Stimuli

The stimuli were presented in static form as simulated conversations with an AI chatbot. They included a generic user prompt requesting information about one of the topics, along with an AI response. Two topics were used for this study:

**Broad: migration in Switzerland.** It covered information about migration in the country in general. The positive (pro-migration) AIR summarized arguments in favor of migration (e.g., economic development, innovation, cultural diversity), the negative (anti-migration) response presented arguments against migration (e.g., rising rent prices, overloaded infrastructure), and the neutral (control) response outlined general facts about migration (e.g., share of different nationalities among immigrants, motives for immigration).

**Specific: anti-migration initiative.** This topic was related to the popular initiative "No to 10 million Swiss!", which proposed limiting immigration to Switzerland to keep the population below 10 million. The initiative was put to a vote on June 14, 2026, and rejected (Der Bundesrat, 2026); however, at the time of data collection (August 2025), it had not yet been scheduled for a vote. In the positive (pro-migration) condition, AIR outlined arguments against the initiative, whereas in the negative (anti-migration) condition, it presented arguments for the initiative. The neutral response in a control condition presented general facts about the initiative.

To create stimuli, we used arguments common in public debates among proponents and opponents of migration and anti-migration initiatives. Stimuli for each topic began with the same sentence and were similar in length and style to ensure comparability (see examples in Figure 2).

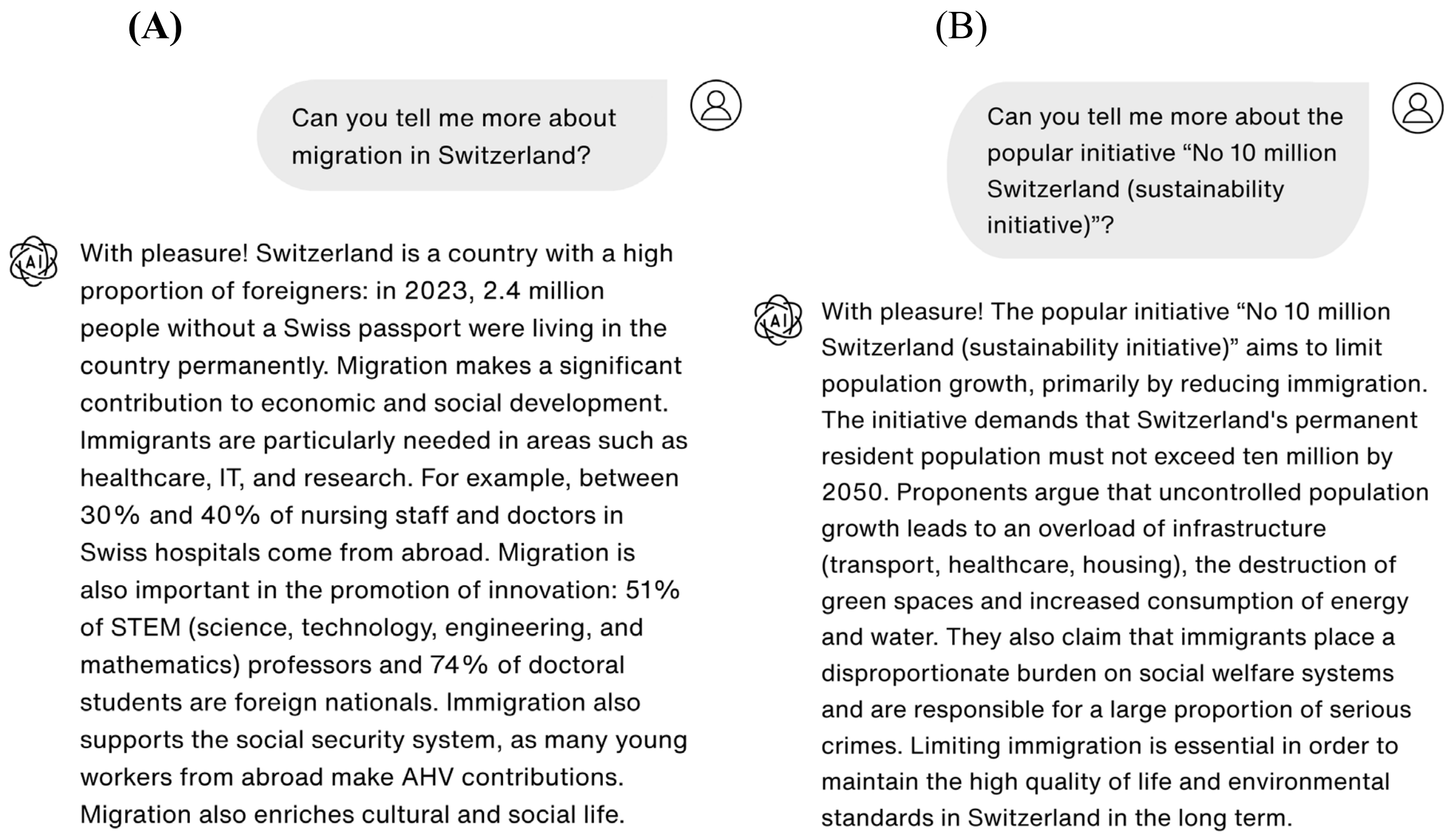

**Figure 2. Stimuli: (A) Positive AIR on migration; (B) Negative AIR on initiative.**

We used ANOVA to evaluate the results of the subjective manipulation check, which, however, indicated that the manipulation was only partially effective. For the broad topic (migration), it was successful, with participants rating negative AIR significantly more negative than neutral AIR and neutral AIR as more negative than positive AIR (all Tukey-corrected pairwise comparisons were significant; $p < .001$). For the specific topic (initiative), however, manipulation was not effective: ratings did not differ significantly between conditions.

## Prompts labeling

The prompts collected as open-ended questions were labeled according to two variables: relevance (i.e., whether a prompt was meaningful and relevant to the task) and prompt stance. The latter was initially divided into three categories: negative (anti-migration), positive (pro-migration), or neutral. Negative prompts were operationalized as those that express a clearly critical or skeptical view of migration or refer to its disadvantages. As positive, we considered prompts that express a benevolent or constructive view of migration or that refer to its advantages. Neutral prompts were those that did not express a specific view and instead inquired about general facts or mentioned both positive and negative aspects of migration.

As we collected an extensive corpus of prompts (8,520 in total before data cleaning), we automatically labeled them. To do so, we used the command-a-03-2025 model from Cohere (2025), which demonstrates high performance across a broad range of tasks. We constructed the prompt based on the operationalization of the categories and included specific examples from the dataset (the full prompt and documentation for the labeling procedure are available in the supplementary material).

To validate the model's performance, a student assistant selected a test dataset of 224 prompts, designed to represent all prompt categories for both topics in a balanced way. The labeling done by the assistant was then checked by the first author, and disagreements were discussed and consensus-coded. To assess its performance, we calculated the standard set of metrics for evaluating computational content analysis approaches: precision, recall, and F1 score (Makhortykh et al., 2024). Our results yielded a macro F1 of 0.92 for relevance and 0.85 for sentiment (per-class F1 0.79–0.97), indicating strong performance.

After irrelevant prompts were removed and some participants were excluded, as described above, the final dataset comprised 4,294 relevant prompts. However, because positive prompts were rare ($N = 322$), they were combined with neutral prompts, and the variable was dichotomized. In the following analysis, we use prompt stance as a binary variable comparing negative prompts (1) against positive/neutral prompts (0), with the latter serving as a baseline. The decision implied a departure from the pre-registered analyses (i.e., a generalized linear mixed model (GLMM) instead of a linear mixed model (LMM); a drop of contrast between neutral and positive prompts and the associated interactions).

## Measures

### *Independent variable*

The key predictor of this study is migration beliefs that were measured using six items adapted from the European Social Survey (European Social Survey European Research Infrastructure (ESS ERIC), 2023): (1) "To what extent do you think Switzerland should allow people of the same race or ethnic group as most people in Switzerland to come and live here?", (2) "How about people of a different race or ethnic group from most people in Switzerland?", (3) "How about people from the poorer countries outside Europe?", (4) "Would you say it is generally bad or good for the economy in Switzerland that people come to live here from other countries?", (5) "Would you say that cultural life in Switzerland is generally undermined or enriched by people coming to live here from other countries?", (6) "Is Switzerland made a worse or a better place to live by people coming to live here from other countries?". Items 1-3 were measured on 4-point Likert scales, whereas items 4-6 were measured on 11-point Likert scales. To create an index variable, the mean of the standardized (z-score) available items was calculated.

The internal consistency among the items was high, with a standardized Cronbach's α of .91.

### *Additional predictors and controls*

As control variables, we used gender, age, and education (having vs. not having a university degree). Additionally, for interaction analysis, we measured participants' attitudes towards AI. Specifically, their *trust in AI chatbots* ("To what extent do you agree or disagree that AI chatbots provide correct and trustworthy information?") and *perceived AI competence* ("Generally speaking, how well do you understand how AI chatbots work?"). Furthermore, we measured participants' thinking styles using a scale developed by Newton et al. (2024) that differentiates among *Actively open-minded thinking (AOT)*, *Closed-minded thinking (CMT)*, *Preference for effortful thinking (PET)*, and *Preference for intuitive thinking (PIT)* (all items available in supplementary materials).

## Data analysis

## Preregistered model

We analyzed the data using binomial generalized linear mixed models (GLMM), with prompt stance as a dependent variable, theoretically relevant predictors as fixed effects, and participant as a random effect. To test the hypotheses, we started with the more complex, preregistered model, using a custom contrast matrix. We began with a mixed model that included all main effects and interactions of Time (pretest/posttest), AIR (positive/negative/neutral), Topic (broad/specific), and both the linear and quadratic migration belief terms, and then selected a subset of theoretically motivated indicator variables.

First, these included main effects and simple interactions of the linear belief slope, Time, and Topic. Second, we included three-way interactions of Time and AIR with both linear and quadratic belief terms. The quadratic term was used exclusively in three-way interactions, based on the assumption that the effect of migration attitudes on prompt stance will be amplified by exposure to AIR. Finally, we included the interactions between Time and AIR reflecting intercept shifts.

We excluded the main effect of the quadratic belief term and the two-way interactions involving it because we lacked the relevant theoretical assumptions. We also excluded the main effects of the AIR condition and its simple interactions with migration beliefs, because any effect of the AIR condition could only manifest at posttest. This resulted in the following model:

$$
\begin{aligned}
\text{logit}\left(\text{prompt}_{ij}\right) = \\
&\beta_0 + \beta_1\text{belief} + \beta_2\text{time} + \beta_3\text{topic} + \beta_4(\text{belief} \times \text{time}) + \beta_5(\text{belief} \times \text{topic}) \\
&\quad + \beta_6(\text{time} \times \text{topic}) \\
&+ \underbrace{\left(\beta_7 + \beta_8\text{belief} + \beta_9\text{belief}^2\right)}_{\text{negative AIR at posttest}} \\
&+ \underbrace{\left(\beta_{10} + \beta_{11}\text{belief} + \beta_{12}\text{belief}^2\right)}_{\text{positive AIR at posttest}} + u_j
\end{aligned}
$$

**Note:** $u_j$ is the random intercept for participant $j$

## Exploratory analysis

After testing the preregistered model, we conducted a full-factorial analysis starting with the most complex model, which included all interactions between migration beliefs (both as linear and quadratic terms), AIR, Time, and Topic. We gradually reduced the model's complexity by removing interactions (see the repository for detailed specifications of each model) and also tested the simplest model with only a linear belief term. Models were compared using the Likelihood Ratio Test. The best fit ($\chi^2(2) = 6.99$, $p = 0.03$; compared to a simpler model) was provided by the model that included interactions between beliefs and topic but used AIR and time only for main effects (M2 in the repository). Subsequent, more complex models did not provide a significantly better fit.

$$
\begin{aligned}
\text{logit}\left(\text{prompt}_{ij}\right) \\
&= \beta_0 + \beta_1\text{belief} + \beta_2\text{belief}^2 + \beta_3\text{topic} + \beta_4(\text{belief} \times \text{topic}) \\
&+ \beta_5\left(\text{belief}^2 \times \text{topic}\right) + \beta_6\text{time} + \beta_7\text{AIR}_{\text{neg}} + \beta_8\text{AIR}_{\text{pos}} + u_j
\end{aligned}
$$

To answer RQ1 and RQ2, we tested whether the main effect of migration beliefs on prompt stance is moderated by other variables, including thinking styles and AI attitudes. For that, we used the model M2 as the basis and added interactions with both linear and quadratic terms. For trust in AI chatbots and perceived AI competence, quadratic models provided a better fit, whereas linear models were better for thinking styles.

# Results

## Descriptives

The descriptive statistics for the variables of interest are displayed in Table 1. In terms of migration beliefs (presented in the table as separate items rather than an index variable for better comprehensibility), we observe that participants have, on average, more positive than negative beliefs, with means on the upper end of the scales. Participants expressed the strongest support for statements suggesting that more people of the same race or ethnic group should be allowed to live in Switzerland and that migration is rather good than bad for the economy. The least supported statements suggested that people from poorer non-European countries should live in Switzerland and that migration in general is rather good than bad.

**Table 1. Descriptive statistics for the independent and control variables**

| Variable | Category/Item | Mean (SD) | Count (Percentage) |
|---|---|---|---|
| Age | | 44.3 (15.1) | |
| Gender | Women | | 279 (49.6%) |
| | Men | | 284 (50.4%) |
| Education | No university degree | | 388 (68.9%) |
| | University degree and higher | | 174 (30.9%) |
| | | | NA = 1 |
| Migration beliefs | Allowing people of the same race or ethnic group into Switzerland (1-4) | 2.7 (0.8) | |
| | Allowing people of a different race or ethnic group into Switzerland (1-4) | 2.5 (0.8) | |
| | Allowing people from the poorer countries outside Europe into Switzerland (1-4) | 2.4 (0.8) | |
| | Immigration bad/good for economy (1-11) | 6.4 (2.6) | |
| | Immigration bad/good for cultural life (1-11) | 5.7 (2.8) | |
| | Immigration bad/good in general (1-11) | 5.6 (2.6) | |
| Preference for effortful thinking (PET) (1-6) | | 4.2 (1.0) | |
| Preference for intuitive thinking (PIT) (1-6) | | 4.0 (0.8) | |
| Actively open-minded thinking (1-6) | | 3.7 (0.9) | |
| Closed-minded thinking (1-6) | | 3.3 (1.0) | |
| Trust in AI chatbots (1-5) | | 3.6 (0.8) | |
| Perceived AI competence (1-5) | | 3.5 (0.9) | |
| Prompt stance | Negative | | 1,322 (30.8%) |
| | Positive/Neutral | | 2,972 (69.2%) |

**Note:** Prompt stance is a prompt-level variable, whereas the rest are participant-level variables

Participants scored the highest on the preference for effortful thinking scale and lowest on the closed-mindedness scale. Furthermore, on average, they considered AI chatbots rather trustworthy and rated their AI competence relatively highly.

In terms of prompts, 30.8% of prompts used in the analysis were coded as negative (anti-migration), whereas 69.2% fell into the positive/neutral category.

## Hypotheses testing

The results of the preregistered model are presented in Table 2. Migration beliefs had a strong negative association with negative prompts ($b = -0.78$, $p < .001$). In other words, the more critical a person was of migration, the more likely they were to formulate a negative prompt. Specifically, each standard deviation increase in migration beliefs was associated with roughly half the log-odds of using a negative prompt. Thus, H1 was supported.

**Table 2. Mixed-effects logistic regression predicting the use of negative prompts**

| **Predictor** | **Log-odds** | **AME (Δ probability)** |
|---|---|---|
| Intercept | -1.279*** | |
| | (0.079) | |
| Migration beliefs | -0.785*** | -0.126*** |
| | (0.096) | (0.015) |
| Time | 0.095 | 0.015 |
| | (0.137) | (0.022) |
| Topic | -0.244 | -0.039 |
| | (0.157) | (0.025) |
| Time × Migration beliefs | -0.245 | -0.039 |
| | (0.169) | (0.027) |
| Topic × Migration beliefs | -0.453** | -0.073** |
| | (0.168) | (0.027) |
| Time × Topic | -0.066 | -0.011 |
| | (0.158) | (0.025) |
| Time × AIR (negative) | 0.414* | 0.069* |
| | (0.187) | (0.033) |
| Time × AIR (negative) × Migration beliefs | 0.282 | 0.045 |
| | (0.220) | (0.035) |
| Time × AIR (negative) × Migration beliefs$^2$ | 0.110 | 0.018 |
| | (0.146) | (0.024) |
| Time × AIR (positive) | 0.213 | 0.035 |
| | (0.204) | (0.034) |
| Time × AIR (positive) × Migration beliefs | 0.189 | 0.030 |
| | (0.215) | (0.035) |
| Time × AIR (positive) × Migration beliefs$^2$ | 0.136 | 0.022 |
| | (0.141) | (0.023) |
| SD (Intercept Subject) | 1.256 | |
| Number of observations | 4294 | |
| R2 Marginal | 0.111 | |
| R2 Conditional | 0.399 | |
| AIC | 4603.4 | |
| BIC | 4692.5 | |

SEs in parentheses. Reference levels: Topic = Specific, Time = Pretest, AIR = AIR (neutral). * p < .05, ** p < .01, *** p < .001

We further observed a significant interaction between the topic and migration beliefs ($b = -0.45$, $p = .007$), suggesting that the effect of beliefs on the use of negative prompts was stronger for a broad topic than for a specific one. This result supports H2.

Exposure to positive or negative AIR, however, did not amplify or mitigate the effect of migration beliefs on the prompt stance: none of the three-way interactions were significant. Therefore, there was no reliable evidence for H3. However, there was a positive effect of negative AIR at posttest on the use of negative prompts ($b = 0.41$, $p = .027$), indicating that participants exposed to an anti-migration AI response were more likely to formulate negative prompts regardless of their individual beliefs.

## Robustness analysis

As discussed in Data Analysis, we gradually reduced the complexity of the preregistered model to achieve a better fit. A better-fitting model included only the interaction between migration beliefs and topic, but not interactions among beliefs, AIR, and time. The fact that adding these interactions did not further improve the model is consistent with the absence of evidence for the effect of AIR on attitude-consistent prompts. Furthermore, in more complex models, neither of the interactions, including both topic and AIR or topic and time, was significant, suggesting that while manipulation failed for the initiative, it likely did not cause a substantial difference in prompt stance at posttest.

The simplified model also supported H1 and H2, with the main effect of migration beliefs and the interaction between beliefs and topic remaining significant (see Figure 3). Additionally, this model shows a positive effect of time on the use of negative prompts ($b = 0.39$, $p < .001$), suggesting that participants were more likely to formulate negative prompts at posttest.

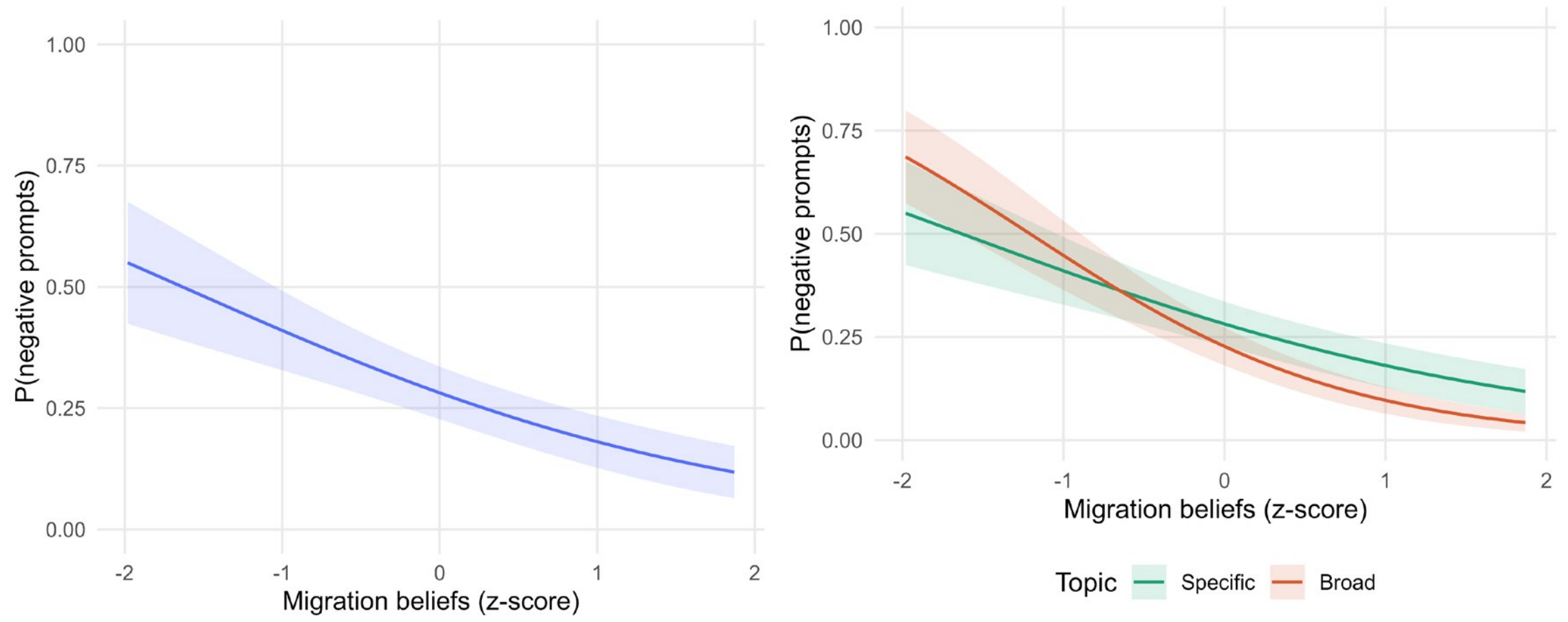

**Figure 3. Probability of using a negative prompt depending on migration beliefs: (A) general, (B) for broad and specific topics.**

When we added control variables—age, gender, and education—to the model (see Model_Controls in the supplementary materials), both the migration beliefs effect (H1) and its interaction with the topic (H2) remained significant. Furthermore, there were main effects of demographics on the outcome variable. In particular, age was positively associated with the prompt stance ($b = 0.03$, $p < .001$), such that older participants were more likely to formulate negative prompts. Having a university degree, in turn, was negatively associated with the use of negative prompts ($b = -0.33$, $p = .025$).

## Interaction analysis

To answer RQ1 and RQ2, we conducted an exploratory analysis of potential moderators of selectivity in prompt formulation. Specifically, we tested whether thinking styles—AOT, CMT, PET, and PIT—or attitudes towards AI—trust and perceived competence—could amplify or mitigate selective exposure. None of the interactions (full models are available in the supplementary material) was significant. Thus, we have no evidence that thinking styles or AI attitudes affected the propensity to formulate attitude-confirming prompts.

We did, however, find a positive effect of AI trust on the use of negative prompts ($b = 0.32$, $p = .004$), indicating that participants who perceived AI chatbots as more trustworthy and credible were more likely to formulate anti-migration prompts.

# Discussion

The introduction of generative AI has been changing how people search for political information online. AI chatbots generate outputs on a variety of topics and engage in human-like interactions with their users. Their features—including their conversational nature, high personalization, and anthropomorphism—lead to dramatic shifts in information seeking. In particular, a user does not merely request or select information but has the opportunity to co-create it through ongoing interaction with a chatbot. Yet user inputs can trigger sycophantic behavior in AI chatbots, as these tools tend to adjust their responses to align with users' political views (Bianchi et al., 2026). This, in turn, may have a considerable impact on both citizens' political decision-making and their awareness of political issues. It is therefore important to understand how users request political information from AI chatbots—specifically, how they phrase their prompts—and whether their interactions reflect their political beliefs.

This study experimentally investigated how Swiss citizens formulate prompts for AI chatbots on migration-related issues. In line with the proposed hypothesis, we found that users tend to construct prompts consistent with their beliefs. Specifically, the more negative participants' attitudes towards migration were, the more likely they were to suggest negative prompts—for instance, explicitly expressing an anti-migration stance or seeking critical information about the topic. This suggests that prompt formulation is affected by selective exposure, which, in the context of AI chatbots, may manifest more through selective prompting than through selective choice of information, which, to date, has been the primary focus of existing scholarship.

We further demonstrate that the effect of beliefs on the prompt stance was stronger for a broad topic (migration) than for a specific one (anti-migration initiative). This finding is consistent with existing research on the preference for congenial information (Hart et al., 2009). While both topics can be considered polarizing and value-relevant, AI response about the initiative likely had greater utility for participants, as the proposal was about to be put to a vote and thus such information was relevant to their voting decisions. This, in turn, might have increased accuracy motivation, thereby reducing the preference for attitude-consistent prompts compared with a broader, more abstract debate on migration. Similar to studies on search queries (van Hoof et al., 2024; Vziatysheva et al., 2026), this demonstrates that selectivity in online search is context-dependent.

However, we did not find an effect of exposure to an attitude-consistent or attitude-discrepant AI response on prompt formulation—it neither amplified nor reduced the selective exposure. This can be partly explained by methodological factors (not sufficiently strong stimuli; static, non-interactive format), but also by the nature of information seeking. If driven by the desire to find specific information, a person might formulate prompts based on their search goals, unrelated to AI outputs, rather than engaging in a back-and-forth conversation with a chatbot and following up on its responses. On the other hand, the effects of exposure to slanted AI outputs should not be completely disregarded. In particular, we observed that exposure to the negative AIR increased the probability of formulating negative prompts, irrespective of one's beliefs, though the opposite was not true for the positive AIR. This suggests that AI outputs can generally drive more ideologically biased prompting, even when it is not aligned with one's beliefs.

Furthermore, we find that trust in AI chatbots increased the probability of using negative prompts. While trust did not moderate selective exposure, this result can indicate that users with higher levels of trust in AI might be more willing to use more expressive (and potentially less socially desirable) prompts. Finally, we did not find any evidence that thinking styles affect selectivity in prompting or prompt formulation in general. Though this should not be interpreted as a lack of effects of cognitive processes on information seeking, this might indicate that different thinking styles drive search request formulation to a lesser extent than, for instance, information perception.

Taken together, our findings contribute to the research on selective exposure in political information seeking and on human-AI interactions. Based on our results, we can suggest several directions for future research. First, it is important to study prompts in the context of actual interactions with chatbots—both experimentally and through analysis of real-life conversations—as they may be more prone to selectivity and attitude-confirming behaviors than in the case of our simulated static AI responses. Second, it is important to compare selectivity in prompting across topics beyond migration. For instance, prompting strategies may depend on the level of information utility, the user's knowledge of the issue, or the issue's relevance. Finally, selectivity in prompting can be affected not only by what AI chatbots communicate but also by how they do so—for example, the tone of the generated language (e.g., more formal vs. more personal), the degree of personalization, and the mode (e.g., textual vs. voice). This prompts an investigation into how specific features of AI chatbots can affect user interactions with them in the context of political issues.

## Limitations

This study has several limitations. First, the experiment did not involve a real human-AI interaction, whereby participants provided prompts in the form of open-ended questions. Therefore, the prompts likely did not fully represent actual user behavior. Since participants were exposed to static stimuli and could not receive answers to their specific questions, they may have paid less attention to AI responses and been less motivated to provide more conversational prompts typically used for AI chatbots. Second, this study did not assess the extent to which participants relied on cues provided in AI responses when formulating prompts. For instance, they could have directly inquired about specific points mentioned in the stimuli, which would indicate a stronger influence of AI responses on subsequent user behavior, or they could have asked questions unrelated to the texts they saw. Third, the manipulation was ineffective for the initiative topic, which again might have contributed to negative results for H3. Finally, this study used a binary measure of prompt stance (partly due to the data quality issues discussed above and partly due to the complexity of more fine-grained coding) rather than a scale, as intended in the preregistration, potentially reducing the statistical power of the models.

## Deviation from preregistration

The main deviation from preregistration concerns the coding of the dependent variable, i.e., prompt stance (binary instead of a 5-point scale), and the subsequent choice of model (binomial GLMM instead of LMM). This decision was driven by the quality of open-ended responses, which often lacked sufficient context to support more fine-grained coding, and by the sparsity of the data (i.e., a small number of positive prompts). However, this possibility was already anticipated in preregistration.

## Disclosure of interest

The authors reports there are no competing interests to declare.

## Data availability statement

The data underlying this article will be shared on reasonable request to the corresponding author.

## Supplementary material

Supplementary materials for this article are available in an online repository: https://osf.io/vb8rz/overview?view_only=e4663cb42cf349aaaf23d91d5e074a87.